\documentclass[twocolumn]{aastex631} 
\usepackage[colorinlistoftodos]{todonotes}
\usepackage{amsmath}
\newcommand{\be}{\begin{eqnarray}}
\newcommand{\ee}{\end{eqnarray}}

\usepackage{CJKutf8}
\usepackage{hyperref}
\usepackage{xcolor}

\begin{document}
	
\title{Stellar Spin Down in Post-Mass Transfer Binary Systems}

\begin{CJK*}{UTF8}{gbsn}

\correspondingauthor{Meng Sun}
\email{meng.sun@northwestern.edu}

\author[0000-0001-9037-6180]{Meng Sun(孙萌)}
\affiliation{Center for Interdisciplinary Exploration and Research in Astrophysics (CIERA), Northwestern University, 1800 Sherman Ave, Evanston, IL 60201, USA}

\author[0000-0001-6692-6410]{Seth Gossage}
\affiliation{Center for Interdisciplinary Exploration and Research in Astrophysics (CIERA), Northwestern University, 1800 Sherman Ave, Evanston, IL 60201, USA}


\author[0000-0002-3944-8406]{Emily M. Leiner}
\affiliation{Department of Physics, Illinois Institute of Technology, Chicago, IL 60616, USA}
\affiliation{Center for Interdisciplinary Exploration and Research in Astrophysics (CIERA), Northwestern University, 1800 Sherman Ave, Evanston, IL 60201, USA}

\author[0000-0002-3881-9332]{Aaron M.\ Geller}
\affiliation{Center for Interdisciplinary Exploration and Research in Astrophysics (CIERA), Northwestern University, 1800 Sherman Ave, Evanston, IL 60201, USA}
\affiliation{Department of Physics \& Astronomy, Northwestern University, 2145 Sheridan Road, Evanston, IL 60208-3112}


\begin{abstract}
Motivated by measurements of the rotation speed of accretor stars in post-mass-transfer (post-MT) systems, we investigate how magnetic braking affects the spin-down of individual stars during binary evolution with the \texttt{MESAbinary} module. Unlike the conventional assumption of tidal synchronization coupled with magnetic braking in binaries, we first calculate whether tides are strong enough to synchronize the orbit. Subsequently, this influences the spin-down of stars and the orbital separation. In this study, we apply four magnetic braking prescriptions to reduce the spin angular momentum of the two stars throughout the entire binary evolution simulation. Our findings reveal that despite magnetic braking causing continuous spin-down of the accretor, when the donor begins to transfer material onto the accretor, the accretor can rapidly spin up to its critical rotation rate. After MT, magnetic braking becomes more important in affecting the angular momentum evolution of the stars. Post-MT accretor stars thus serve as a valuable testbed for observing how the magnetic braking prescriptions operate in spinning down stars from their critical rotation, including the saturation regimes of the magnetic braking. The rotation rate of the accretor star, combined with its mass, could provide age information since the cessation of MT. By comparing the models against observation, the magnetic braking prescription by \citet{2018ApJ...862...90G} is found to better align with the rotation data of post-MT accretors.
\end{abstract}


\section{Introduction}
\label{sec:intro}

When constructing a stellar evolutionary model, mass, metallicity, rotation, stellar wind, and microphysics describing energy transportation are considered as crucial initial inputs. However, stellar rotation, angular momentum evolution, as well as magnetic fields, are frequently ignored in the initial simplification steps due to the presence of numerous unresolved questions associated with their related physics. Observationally, both {\it Kepler} and its successor, {\it TESS} mission, reveal rotation information for main sequence (MS) stars and evolved late-type stars \citep{2013MNRAS.432.1203M,2013ApJ...775L..11M,2013A&A...557L..10N,2013MNRAS.436.1883W,2014ApJS..211...24M,2014A&A...572A..34G,2015A&A...583A..65R,2016MNRAS.456..119C,2017A&A...605A.111C,2019ApJS..244...21S,2020AJ....160..168L,2020ApJS..250...20C,2021AJ....161..189L,2021ApJS..255...17S,2022ApJ...930....7A,2023A&A...678A..24R}, with a focus on field single stars. Stellar rotation measurements are also conducted in star clusters using both ground-based and space telescopes, covering a wide range of cluster ages \citep{2008MNRAS.384..675I, 2011ApJ...733..115M, 2011ApJ...733L...9M, 2012ApJ...747...51H, 2013MNRAS.430.1433A, 2013A&A...560A..13M, 2015Natur.517..589M, 2015ApJ...809..161N, 2016ApJ...823...16B, 2017ApJ...842...83D, 2017ApJ...839...92R, 2018AJ....155..196R, 2020ApJ...904..140C, 2020A&A...641A..51F, 2021ApJS..257...46G}, allowing for further studies of single star gyrochronology, which is a method to obtain age information by measuring the stellar rotation rate.

However, studying stellar rotation in a binary system is less discussed, typically requiring long-term spectroscopy to fully characterise the orbit. Most mass transfer (MT) binary systems that contain at least one unevolved star can have orbital periods ranging from days to years, necessitating extended observations to capture multiple phases of the stellar orbit. A recent systematic study of blue straggler and blue lurker stars provides such an opportunity. Blue straggler stars (BSSs) are most easily defined in star clusters as stars that are more luminous or hotter than the MS turn-off on the color-magnitude diagram. In open clusters, as the stellar density is not high enough to allow for frequent mergers or direct collision events, the formation of BSSs is believed to occur primarily through a MT phase \citep{2009Natur.462.1032M, 2011Natur.478..356G, 2015ApJ...814..163G}. The donor star loses its envelope, becoming a white dwarf (WD), while the accretor star gains mass, evolving beyond the MS turn-off and ultimately becoming a BSS. Blue lurker stars are believed to be the low-mass counterparts of BSSs, thought to also form through MT. Blue lurkers are still located among the MS stars on the color-magnitude diagram, appearing as fast rotators that cannot be explained through single-star evolution \citep{2019ApJ...881...47L}. 

Several observational studies have documented that fast rotation is a frequent outcome of BSS and blue lurker formation. A recent study by \citet{2023NatCo..14.2584F} shows that fast-rotating BSSs in globular clusters prefer low-density environments. This suggests that post-merger or collision products may have a shorter spin-down timescale than expected. Other studies by \citet{2018ApJ...869L..29L} and \citet{2023ApJ...944..145N} collected BSSs from open clusters and a few field binaries containing a low-mass MS star and a WD. They provide the rotation period of the BSSs and MS stars as a function of the cooling age measured from their WD companion. The cooling age of the WD can be considered as the system age after MT stops. Since MT accelerates the rotation of the accretor stars, this process acts like resetting the gyro-clock, allowing observation of how magnetic braking subsequently slows down those rapid rotators.

A robust theory of magnetic braking has been developed for single main-sequence stars, describing how the magnetized corona wind escapes from the surface of stars and reduces their angular momentum. This research began more than sixty years ago \citep{1962AnAp...25...18S,1967ApJ...148..217W, 1968MNRAS.138..359M}. The spinning down in low-mass stars, specifically stars with a convective outer envelope, has been confirmed by observation around the same time \citep{1967ApJ...150..551K, 1972ApJ...171..565S}. However open questions still persist to this day, and challenges arise, such as the stalled spin-down phenomenon observed in stars with a thicker convective envelope ($<0.8$ $M_{\odot}$), occurring between a star age of 0.7 to 1 Gyr \citep{2019ApJ...879...49C, 2020ApJ...904..140C, 2022ApJ...929..169R}. Possible solutions could involve the saturation of magnetic braking efficiency due to a relatively complex surface magnetic field, or the treatment of core and envelope rotation as separate issues \citep{1991ApJ...376..204M, 1995ApJ...441..865C, 1995A&A...294..469K, 1998A&A...333..629A, 2000ApJ...534..335S, 2005A&A...440..973V, 2011MNRAS.416..447S, 2015ApJ...799L..23M, 2015A&A...584A..30L,2015ApJ...813...40G,2016A&A...595A.110G,2020A&A...636A..76S, 2023ApJ...951L..49C}.


Here, we seek to extend the magnetic braking theory to low-mass binary systems by developing the first 1-D binary evolutionary simulation that implements four possible magnetic braking prescriptions in wide, non-synchronized low-mass binary systems. This is in contrast to the traditional idea that strong tidal action accompany magnetic braking, where angular momentum is removed directly from the orbit instead of from the stars, as the stellar spin is always assumed to be synchronized with the orbit \citep{2015ApJS..220...15P}. We use the binary module of the Modules for Experiments in Stellar Astrophysics (\texttt{MESA}, version 11701; \citealt{2011ApJS..192....3P, 2013ApJS..208....4P, 2015ApJS..220...15P, 2018ApJS..234...34P, 2019ApJS..243...10P, 2023ApJS..265...15J}). Besides the \texttt{MESA} code, the infrastructure of the next-generation binary population synthesis code \texttt{POSYDON} (POpulation SYnthesis with Detailed Binary-evolution simulatiONs, \citealt{2023ApJS..264...45F}), is used to launch a multitude of binary grids covering a vast parameter space. We briefly introduce the observational data used to validate our model in Section \ref{sec:obs}. In Section \ref{sec:MB in binaries}, we present a more realistic approach to applying magnetic braking in detailed binary modeling. In Section \ref{sec:POSYDON-MESA grid}, we list the important physics in setting up the grid. We provide the results from the default binary grid in Section \ref{sec:result}, and further comparisons of the additional magnetic braking theories are presented in Section \ref{sec:other MB}. Potential physics that could have a slight impact on the results, along with a discussion of the promising magnetic braking prescription, are presented in Section \ref{sec:Discussion}. We conclude with our main findings in Section \ref{conclusion}.

\section{Observations}
\label{sec:obs}

\citet{2018ApJ...869L..29L} includes 12 post-MT systems from the field and open clusters \citep{2015ApJ...814..163G,2021AJ....161..190G} containing one FGK-type MS star and a WD. The presence of a white dwarf companion and an orbital period ranging from 100 to 3000 days provides direct evidence that the system has evolved through MT, and WD cooling models can provide the time since this mass transfer occurred in the system. The systems' orbital separations are close enough to ensure that the system has experienced an MT phase, but wide enough that the rapid rotation observed in the MS stars cannot be explained by tides.  The hypothesis is that the accretor spins up to near break up during MT, and then spins down following standard gyrochronology models for low-mass MS stars. Therefore, gyrochronology may be used in post-MT binaries to measure the age of the system since MT occurred. 

In open clusters, BSSs are believed to be formed mainly through MT \citep{2009Natur.462.1032M,2011Natur.478..356G}, and thus may by expected to rotate rapidly shortly after MT formation. \citet{2019ApJ...881...47L} identified rapidly rotating stars among the MS stars on the color-magnitude diagram of M67, suggesting these may be hidden post-MT stars with lower masses than the blue stragglers. These stars are named as blue lurkers. Using the {\it Hubble Space Telescope} UV spectroscopy, \citet{2023ApJ...944..145N} confirmed the detection of the WD orbiting around two blue lurkers, confirming they formed via MT. This study also measured WD temperatures and cooling ages, thereby adding two more data points in the rotation period versus age parameter space for us to check the magnetic braking models. The compared data are listed in Table \ref{tab:obs}. In the table, systems with a WOCS id indicate that they are from an open cluster. The acronym WOCS stands for the WIYN Open Cluster Survey, which conducted a spectroscopy survey over 20 years using the WIYN 3.5m telescope and has identified numerous BSSs in binaries through the radial velocity method \citep{2000ASPC..198..517M}.

\begin{table*}[]
\caption{Post-Mass-Transfer System Properties from \citet{2018ApJ...869L..29L}, \citet{2023ApJ...944..145N}   and the reference therein}
\centering
\begin{center}
\begin{tabular}{c c c c c c c c}
\hline\hline
ID & Gaia ID	& WD $T_\text{eff}$	& WD $\log\,g$  & WD Age & BS $T_\text{eff}$ &    $P_{\text{rot}}$ &$P_{\text{orb}}$\\
 & & (K) &$(g/{\rm cm\,s^{-2}})$  & (Myr) & (K)& (day) & (day)\\
\hline
WOCS 5379 & 573939679517550848 & $15400^{+280}_{-250}$ & $7.5^{+0.06}_{-0.05}$ & $230^{+40}_{-30}$ & $6400\pm120$  & $>$ 2.5 & 120\\
WOCS 4540 & 573944111923749760  & $17100^{+150}_{-100}$ & $7.7^{+0.04}_{-0.02}$ & $95^{+7}_{-5}$ & $6590\pm100$  & $1.8^{+2.3}_{-0.92}$ & $3030 \pm 70$\\
WOCS 4348 & 573941844181066368 & $13000\pm 500$ & 7.8 & $245^{+30}_{-25}$ & $6750\pm120$ &$1.2^{+1.5}_{-0.6}$&  $1168\pm 8$ \\
WOCS 5350 & 573939954395446016 & $13200\pm 500$ & 7.8 & $235^{+30}_{-25}$ & $6720\pm120$ & $5.3^{+7.1}_{-2.9}$ & $690\pm 3$ \\
WOCS 1888 & 573973111543073024 & $11200\pm 500$ & 7.8 & $370^{+50}_{-40}$ &$6570\pm120$  & $3.6^{+4.7}_{-1.9}$ &  $2240\pm 30$  \\
WOCS 2679 & 573998091072694912 & $11300\pm 500$ & 7.8 & $360^{+50}_{-40}$ & $6630\pm120$ & $1.4^{+1.9}_{-0.8}$ &  $1033\pm 8$   \\
WOCS 4230 & 573944283722429952 & $11800\pm 500$ & 7.8 & $320^{+40}_{-35}$ &$6350\pm110$  & $1.0^{+1.3}_{-0.6}$ &  --  \\ 
WOCS 3001 & 573949914422260736 & 10300 -- 10500 & 7.4-- 7.6& 600 -- 900 &$6690^{+80}_{-160}$ & $2.0^{+0.6}_{-0.9}$ &  128.14  \\
WOCS 14020 & 604906531159503616 & 11000 -- 12200 & 7.0-- 7.6& 300 -- 540 &$5990^{+60}_{-110}$    & 4.4 &  358.9  \\
RE 0044+09$^{a}$ & 2750199780197751936 & $28700\pm 1500$ & 8.41 & $51^{+13}_{-12}$ &  --  & 0.4 &   $>30$ \\
KOI-3278$^{b}$ & 2052729252051464192 & $9960\pm730$ & 8.14 & $663 \pm 60$ & $5568^{+40}_{-38}$  & $12.5\pm 0.1$ &  $88.181^{+0.00025}_{-0.00027}$  \\
KIC 6233093$^{c}$ & 2075331706432389504 & $<10000$ & 8.0 & $>1000$ &    & 17.1 &    \\
2RE J0357+283$^{d}$ & 166865496701799168 & $35000\pm 5000$ & 8.0 & $6.3^{+2.9}_{-2.3}$ & & 0.4 &    \\
HD 217411$^{e}$ & 2610649828824580096 & $37200\pm 300$ & 7.8 & $4.8\pm0.12$ &   & 0.6 &    \\
\hline

\multicolumn{8}{l}{
\begin{minipage}{18 cm}

$^{a}$ Note: RE 0044+09 contains a K dwarf with a WD from either a wide binary or common proper motion pair \citep{1995ApJ...438..364K}.

$^{b}$ Note: KOI-3278 is a self-lensing binary system comprising a G-dwarf and a white dwarf \citep{2014Sci...344..275K}. The mass and radius of the G-dwarf are 1.04 $M_{\odot}$ and 0.96 $R_{\odot}$, respectively.

$^{c}$ Note: KIC 6233093 is another self-lensing binary system containing a white dwarf \citep{2018AJ....155..144K}. The mass and radius of the primary are 1.10 $M_{\odot}$ and 1.88 $R_{\odot}$, respectively.

$^{d}$ Note: 2RE J0357+283 has a rapid rotating K-dwarf and a white dwarf \citep{1996MNRAS.279..180J}.

$^{e}$ Note: HD 217411 is a triple system with a G, K and a white dwarf \citep{2014MNRAS.444.2022H}.
\end{minipage}
}

\end{tabular}
\end{center}
\label{tab:obs}
\end{table*}

\section{Magnetic Braking in binary evolution}
\label{sec:MB in binaries}

\subsection{The Assumption of Orbital Synchronization}

In the context of single star evolution, magnetic braking plays a crucial role in altering the rotation rates from late F to M dwarfs, characterized by strong surface magnetic fields and convective envelopes. The magnetic braking theory originally developed based on observational evidence, where young solar-like stars were generally observed to rotate more rapidly, while older solar-like stars rotated slowly \citep{1972ApJ...171..565S}. The magnetized corona wind, which follows along the open magnetic field lines near the surface of stars, could reduce the star's angular momentum, causing a decrease in the star's spin rate \citep{1962AnAp...25...18S}.

In extremely close low-mass binary systems, magnetic braking can be the primary mechanism for changing the orbital angular momentum. For example, in a system containing a solar-type primary star and a companion star in a 1-day orbit, tides can synchronize the spin frequencies of both stars with the orbital frequency \citep{2006ApJ...653..621M}. If the primary star's spin rate slows down due to magnetic braking, the system must shrink in separation (causing the primary star to spin up again) to maintain synchronization. This is how magnetic braking affects the binary separation. The crucial point here is that a system, always undergoing strong tidal interactions, tends towards spin-orbital synchronization, regardless of binary separation. When magnetic braking is active, angular momentum is extracted directly from the orbit, without affecting the rotation rate of the stars. This assumption is implemented into the \texttt{MESAbinary} module when applying magnetic braking physics to binary evolution, by default. Following the third \texttt{MESA} instrument paper \citep{2015ApJS..220...15P}, the rate of change in orbital angular momentum is written as:
\be
\dot{J}_{\rm orb} = \dot{J}_{\rm gr,\,df}+\dot{J}_{\rm ml,\,df}+\dot{J}_{\rm mb,\,df}+\dot{J}_{\rm ls,\,df}
\ee
where $\dot{J}_{\rm gr,\,df}$, $\dot{J}_{\rm ml,\,df}$, $\dot{J}_{\rm mb,\,df}$, and $\dot{J}_{\rm ls,\,df}$ represents the orbital angular momentum changes due to gravitational wave radiation, mass escaping from the system, magnetic braking, and spin-orbit coupling due to stellar tides, calculated using default \texttt{MESAbinary} code, respectively.

In our calculation, the changes in the system's orbital angular momentum due to magnetic braking and tides are modified in a more realistic manner. In wider orbits, 
tides no longer have the capability to synchronize the system. Throughout this manuscript, orbital synchronization is no longer assumed to be universally true. Our prescription is equivalent to
\be
\dot{J}_{\rm mb,\,df} = 0.
\label{eq:dfMB}
\ee

Recalling the \texttt{MESA} default calculation of the spin-orbit coupling term \citep{2015ApJS..220...15P}
\be
\dot{J}_{\rm ls,\,df} = -\frac{1}{\Delta t}\bigg( \Delta S_1 - S_{1,\mathrm{ml}}\frac{\dot{M}_{1,\mathrm w}}{\dot{M}_{1}}+\Delta S_2 - S_{2,\mathrm{ml}}\bigg),
\ee
the subscripts 1 and 2 represent the donor and accretor stars, respectively. $\Delta t$ is the binary evolution time step, $\Delta S_1$ and $\Delta S_2$ are the spin angular momentum changes of the donor star and the accretor star. The mass loss rate of the donor star, $\dot{M}_1$, is contributed by both the donor's wind loss rate, $\dot{M}_{1,\mathrm{w}}$, and Roche-lobe overflow (RLOF) rate. In cases where there is no RLOF in binaries, meaning $\dot{M}_{1,\mathrm w} = \dot{M}_1$, the spin angular momentum changes from both stars, $S_{1,\mathrm{ml}}$ and $S_{2,\mathrm{ml}}$ , are only contributed by the stellar wind.

$\dot{J}_{\rm ls,\,df}$ then is substituted as below in our calculation:
\be
\begin{split}
\dot{J}_{\rm ls}  = & -\frac{1}{\Delta t}\bigg(  \Delta S_1 - S_{1,\mathrm{ml}}\frac{\dot{M}_{1,\mathrm w}}{\dot{M}_{1}} - S_{1,\mathrm{mb}} \\
& +\Delta S_2 - S_{2,\mathrm{ml}} - S_{2,\mathrm{mb}}\bigg),
\end{split}
\label{eq:otherls}
\ee
where the spin angular momentum of both the donor and accretor stars, denoted as $S_{1,\mathrm{mb}}$, $S_{2,\mathrm{mb}}$ respectively, is reduced by magnetic braking. To implement tidal interaction into our model, we adapted the formulation proposed by \citet{1981A&A....99..126H} and considered differentially rotating stars, as described in Equation 20 of \citet{2015ApJS..220...15P}.

For the initial rotation speed, two stars with zero rotation are assumed in the fiducial grid. This means the binary system is not synchronized at the beginning. 
As the spin-up by accretion for the accretor star is very efficient, the initial rotation status of both stars would not significantly affect the result. In other words, the initial star spin information is erased during the MT. The systems are far from being synchronized, with $P_{\rm rot} \sim 10^{-4} P_{\rm orb}$ when the accretor stars reach the critical rotation rate in many of the models. Using the default magnetic braking prescription, we tested another smaller grid with an initially synchronized orbit instead of two non-rotating stars. The resulting rotation period distribution of the accretor star after MT is the same as the fiducial grid.

\subsection{Four Magnetic Braking Prescriptions}

For the fiducial model analysis in this work, we applied the \citet{2018ApJ...862...90G} magnetic braking prescription. This model has been a promising model when investigating the rotation distribution of single stars in young open clusters and orbital evolution in low-mass X-ray binaries \citep{2023ApJ...950...27G}. For other magnetic braking prescriptions, we refer the reader to Section 3 of \citet{2023ApJ...950...27G} for detailed discussion. Here, we summarize the key differences among those state-of-the-art prescriptions.

The efficiency of magnetic braking in slowing down a star is linked to the star's surface magnetic field, specifically its magnetic field topologies. Two crucial concepts associated with magnetic braking are the \textit{Alfv\'en radius} and the \textit{Rossby number}. The Alfv\'en radius is characterized by the ratio between the magnetic energy and the kinetic energy of the wind. Outside this Alfv\'en radius, the wind is dominated by kinetic energy, and the material no longer co-rotates with the star. Therefore, the amount of angular momentum loss due to magnetic braking is directly related to how the Alfv\'en radius is parameterized — a function of the magnetic field properties and stellar structure information. The Rossby number, $R_o$ , is defined as $R_o = P_{\rm rot}/\tau_{\rm conv}$. And $\tau_{\rm conv}$ is the eddy turnover time at the top of the outer convection zone. The magnetic field dynamo number is inversely proportional to the square of $R_o$. Thus, a small $R_o$, either from a fast rotator or a deep surface convection zone resulting in a large $\tau_{\rm conv}$, implies a strong magnetic field. However, magnetic field activity saturates if $R_o$ drops below a critical Rossby number. This saturation results in insufficient magnetic braking for small $R_o$. Around the concepts of the Alfv\'en radius and a saturated regime, four magnetic braking prescriptions are discussed in detail from 3.3.1 to 3.3.4 in \citet{2023ApJ...950...27G}, including \citet[hereafter G18]{2018ApJ...862...90G}, \citet[hereafter M15]{2015ApJ...799L..23M}, \citet[hereafter CARB]{2019ApJ...886L..31V}, and \citet[hereafter RVJ]{1983ApJ...275..713R}.

Another key point to note is that the prescriptions by M15 and G18 consider saturated magnetic braking and are better matched to observations for single stars. For close binaries with orbital synchronization, where \citet{2023ApJ...950...27G} compares the model with low-mass X-ray binary and ultra-compact X-ray binary data, G18, CARB, and RVJ better match the data.

\section{\texttt{POSYDON} - \texttt{MESA} Low-mass Binary Grid Setup}
\label{sec:POSYDON-MESA grid}

All the detailed binary modeling starts with two stars being born and evolving simultaneously. 
Assuming stars are spherically symmetric, they are modeled in 1-dimensional stellar structures, applying proper physics to describe how energy is generated due to nuclear reactions at the core and shell, as well as how energy is transported from the nuclear burning region to the surface of the stars. Following the \texttt{POSYDON} version 1 grid with solar metallicity $Z_{\odot} = 0.0142$ \citep{2009ARA&A..47..481A}, we apply the same microphysics and macrophysics. For details such as opacity tables, nuclear reaction rates, prescriptions for stellar winds, and factors related to convective and mixing processes, readers are referred to the first \texttt{POSYDON} instrument paper by \citet{2023ApJS..264...45F}.

In terms of the implemented physics related to binary evolution, although most settings still follow to Section 4 of \citet{2023ApJS..264...45F}, here is a brief review of the important physics. These settings could significantly impact the results of the fiducial binary grid.

During the MT phase, angular momentum is transferred onto the accretor star, causing it to spin up. This prescription is applied with the default \texttt{POSYDON-MESA} assumptions. Two scenarios could occur: either an accretion disk is formed (when the impact parameter of the incoming accretion stream is greater than the star's radius) or the material directly impacts the star (when the impact parameter of the incoming accretion stream is less than the star's radius). For these two cases, the specific angular momentum of the impact stream transferred onto the accretor star is calculated differently \citep{1975ApJ...198..383L, 2013ApJ...764..166D}. This angular momentum transfer is highly efficient in both disk and no-disk scenarios, as demonstrated in several previous studies \citep{2013ApJ...764..166D, 2022A&A...667A.122S}, as well as in \texttt{POSYDON} massive binaries \citep{2024arXiv240307172A}.

When a star rotates near its critical rotation rate, an enhanced wind develops, and the wind speed is described in Equation 1 of \citet{2023ApJS..264...45F}. We refer to this process as \textit{rotation limited accretion} throughout this manuscript. This mechanism ensures the star maintains a sub-critical rotation. In most of the low-mass binary grids, if systems undergo a stable MT phase, the accretor stars typically rotate near the critical rotation rate. The enhanced wind counteracts the mass accretion, resulting in the mass accreted onto the accretor star being usually less than 5\% of its total mass. Although a conservative RLOF is assumed, material could still leave the system due to this rotation-driven accretion limit.


Motivated by the observation in \citet{2018ApJ...869L..29L}, which consist of BSSs and WDs, the parameter space is focused on low-mass stars undergoing stable MT phases. Table \ref{table: initial model} presents the ranges and steps of the binary grid. In this paper, physical quantities of the donor star are denoted with the subscript ``1'', and the accretor star has the subscript ``2''. The donor star always evolves first, so the mass ratio between the donor and accretor, $q=M_2/M_1$, is always less than 1.

\begin{table*}[]
\caption{Initial Parameters and Steps of the Binary Grids}
\begin{center}
\begin{tabular}{c c c c c c}
\hline\hline 
$M_1/M_{\odot}$	& $\Delta \log(M_1/M_{\odot})$ & $q=M_2/M_1$ & $\Delta q$ & $P_{\rm orb}/{\rm day}$ & $\Delta \log (P_{\rm orb}/{\rm day})$	\\
\hline
0.8 - 2 & 0.021 & 0.5 - 0.99 & 0.026 & 1 - $10^3$ & 0.16 \\
\hline
\label{table: initial model}
\end{tabular}
\end{center}
\end{table*}

Our inlists and extra subroutines are shared at zenodo \url{https://zenodo.org/records/11081261}.

\section{Results}
\label{sec:result}

In our fiducial grid, for most post-MT systems, the accretor star always spins down starting from the critical rotation rate at the end of the MT. In other words, for most of our systems, magnetic braking does not cause the star to spin down within a short timescale during the MT phase. This study also explores how different magnetic braking mechanisms affect the star when it is near the critical rotation rate.

\begin{figure}[tp]
\includegraphics[scale=0.5,angle=0]{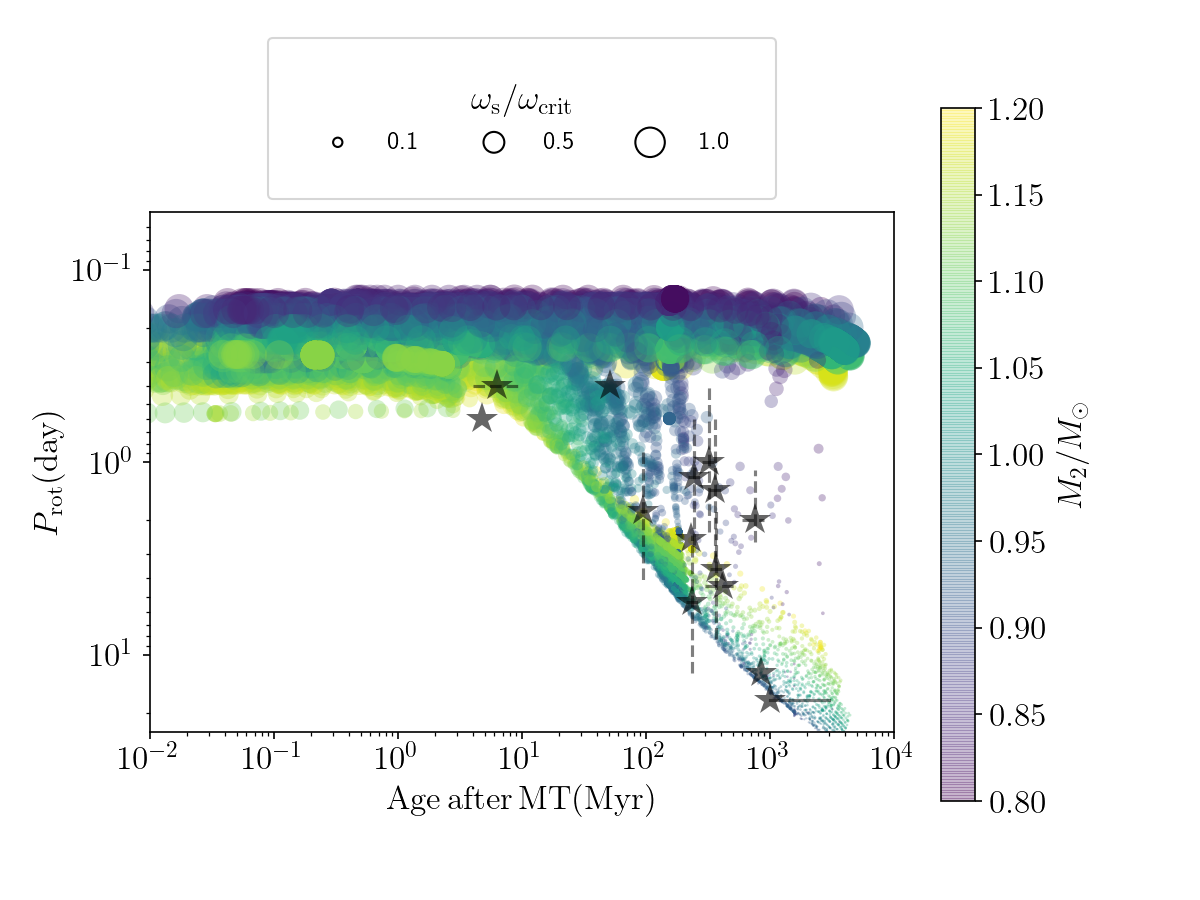}
\caption{The rotation period of the accretor star as a function of the system age after MT stops, using G18's magnetic braking prescription. The size of the dot represents the ratio between the surface rotation frequency and its critical rotation frequency. The colors indicate the mass of the accretor star. Post-MT accretor stars' data are represented by black stars, with error bars shown as black dashed lines.}
\label{fig:Prot_age_fiducial_full}
\end{figure}

In Figure \ref{fig:Prot_age_fiducial_full}, we present the rotation period as a function of the age since MT stops, applied with G18's magnetic braking prescription derived from 3D MHD simulations. The cessation of MT is defined as the MT rate through RLOF drops below $10^{-80}$ $M_{\odot}/{\rm yr}$\footnote{The RLOF MT rate is indicated by the column title `\texttt{lg\_mtransfer\_rate}' from the \texttt{MESA} \texttt{binary\_history.data} file, and we noticed that no MT phase in \texttt{MESA} has $10^{-99}$ $M_{\odot}/{\rm yr}$ in `\texttt{lg\_mtransfer\_rate}'. While $10^{-80}$ $M_{\odot}/{\rm yr}$ and $10^{-99}$ $M_{\odot}/{\rm yr}$ occur at almost the same age.}. The color of the dots represents the accretor's mass during the evolution, and the size of the dot indicates the ratio between the surface rotation frequency $\omega_{\rm s}$ and the critical rotation frequency $\omega_{\rm crit}$ of the accretor star, where the latter term is defined in Equation 1 from \citet{2023ApJS..264...45F}. The data, including error bars from observations, are represented by black stars and dashed lines.

To qualify for inclusion in all the scattered plot throughout the entire manuscript, the system must satisfy the following criteria:

\begin{itemize}
\item The system undergoes stable MT, and the code converges at the end of the simulation.
\item The donor star mass is less than 1.4 $M_{\odot}$, the donor star radius is less than 0.2 $R_{\odot}$, and the donor reaches central H depletion, where the central hydrogen mass fraction (at $r=0$) is below $10^{-4}$, resulting in a WD.
\item The accretor star mass falls between 0.8 to 1.2 $M_{\odot}$,  and these stars have convective envelopes without deep convective envelopes, allowing magnetic braking to significantly affect their rotation.
\item The accretor star central hydrogen mass fraction is greater than $10^{-4}$, indicating it is still a MS star.
\item The orbital period after MT falls between 100 to 5000 days.
\end{itemize}

In general, the models with G18's magnetic braking prescription agree well with the observations. There are two groups in Figure \ref{fig:Prot_age_fiducial_full}. The first group contains accretor stars that haven't been spun down by magnetic braking and maintain a fast rotation rate with $P_{\rm rot} \sim 0.2$ days from 0 to 6 Gyrs since MT stops. The second group contains accretor stars that, after going through stable MT, significantly spun down by magnetic braking from 0 to 0.2 Gyr after MT ceased. After 0.2 Gyr, these stars spin down due to the stellar expansion of the post MS evolution, where $\omega_{\rm s}/\omega_{\rm crit}<0.1$.

\begin{figure}
    \centering
    \includegraphics[scale=0.5,angle=0]{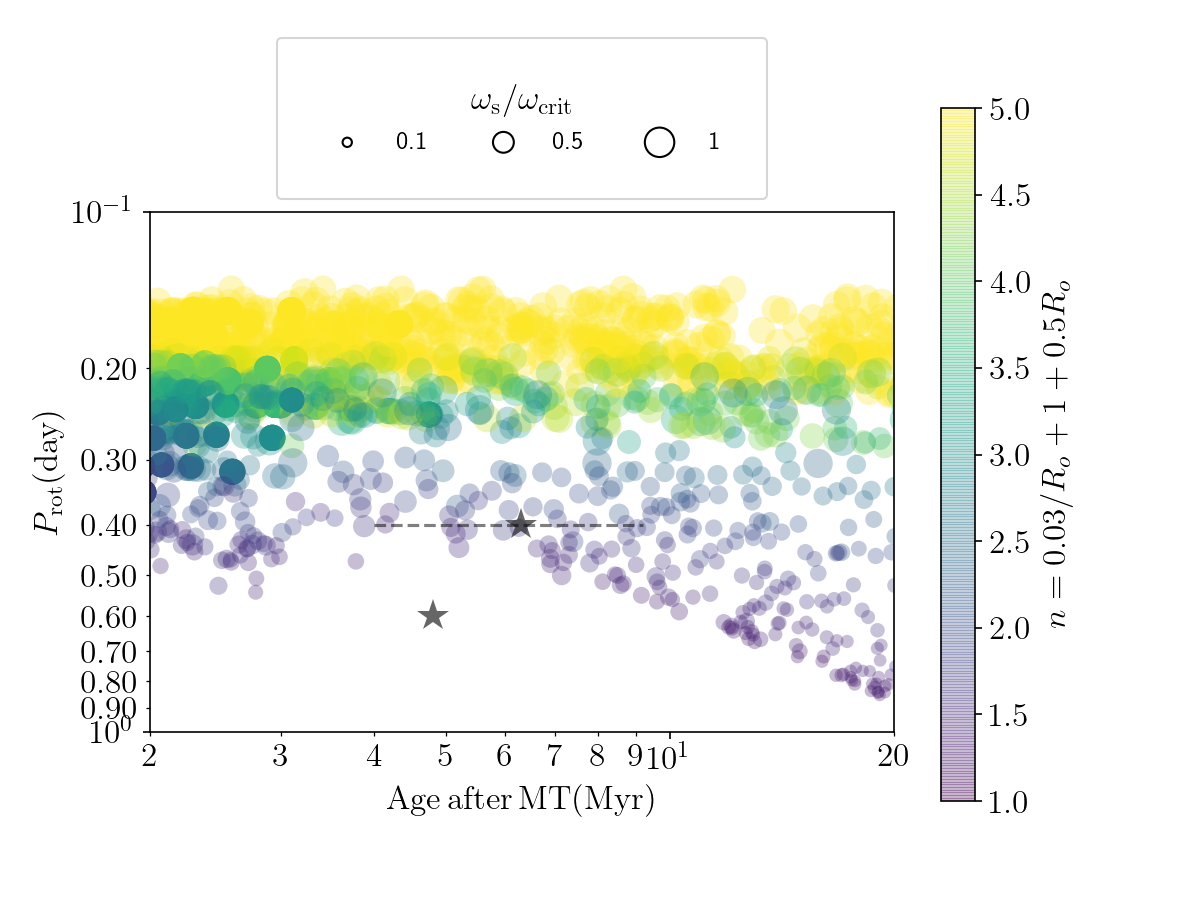}
    \caption{A zoomed-in version of Figure \ref{fig:Prot_age_fiducial_full}, focusing on the region where two groups of models are split. The color of the dots indicates a different physical quantity, describing the complexity of the surface magnetic field of the accretor star.}
    \label{fig:Prot_age_model_g18_full_abn_mgnf}
\end{figure}

The origins of the two groups is shown in Figure \ref{fig:Prot_age_model_g18_full_abn_mgnf}, which is a zoom-in version of Figure \ref{fig:Prot_age_fiducial_full}, specifically focusing on the time near the separation of the two branches of models. The color represents the complexity of the magnetic field $n$ ($n=1$ is the dipole magnetic field and $n>1$ is the high order magnetic field) as described in detail in \citet{2018ApJ...862...90G}:

\begin{equation}
    n = \frac{a}{R_o}+1+b R_o.
\label{equ:n}
\end{equation}

In Figure \ref{fig:Prot_age_model_g18_full_abn_mgnf}, we adopt $a=0.03$ and $b=0.5$, calibrated with open cluster data \citep{2021ApJ...912...65G, 2023ApJ...950...27G}. The two groups split where $n=2$ and $n=3$ in the saturation regime ($R_o \ll 1$). Models on the top with yellow and green color are near the critical rotation rate. Models represented by blue and purple that are lower than the critical rotation rate, could experience significant spin-down due to magnetic braking.

We observe a dense population of post-MT binaries in Figure \ref{fig:Prot_age_fiducial_full}. In the following subsection, we choose a selected number of binary models for detailed analysis with the G18's magnetic braking prescriptions.

\subsection{G18's Magnetic Braking Prescription in Binary Evolution with Conservative MT}
\label{subsec:G18 default result}

The selected binary systems have their initial and final information described in Table \ref{table: selected binaries}. Applying the rotation-limited accretion regime, on average, the accretor stars only gain 0.045 $M_\odot$, while the donor stars lose 0.75 $M_\odot$ and subsequently become helium or carbon-oxygen core WDs. For the systems listed in Table \ref{table: selected binaries}, those with initial $P_{\rm orb} \lesssim 500$ days experience case B mass transfers, initiating during the donor's RGB phase. Conversely, systems with initial $P_{\rm orb} \gtrsim 500$ days undergo case C mass transfers, with stable RLOF commencing during the donor's AGB phase. 

All the accretors reach their critical rotation speed ($\omega/\omega_{\rm crit} \sim 1$) during stable RLOF. While magnetic braking operates to spin down individual stars throughout the entire modeling process, it is not strong enough to cause the accretor star to maintain a sub-critical rotation. Consequently, it would not be able to receive a significant amount of material in most of our binaries. Accretor stars still require an enhanced wind to maintain a sub-critical rotation. 
In most of our low-mass binary evolution models, the final $P_{\rm orb,f}$ is always greater than the initial $P_{\rm orb,i}$. Moreover, within this $P_{\rm orb}$ range, tides play a minimal role in achieving synchronization, and the rotational periods of the stars $P_{\rm rot}$, are much shorter than the orbital periods ($P_{\rm rot} \ll P_{\rm orb}$). In contrast to the traditional assumption of tidal synchronization associated with magnetic braking leading to orbital shrinkage, our system consistently undergoes expansion.

Systems with a donor mass less than 1.02 $M_\odot$ terminate when the system age exceeds the Hubble time, while other systems stop evolving when the donor star becomes a WD. We follow the definition of a WD provided by \citet{2016ApJ...823..102C}, where the central degeneracy parameter is set to be over 10.

\begin{table*}[]
\caption{Information of the Selected Binaries with G18's magnetic braking Prescription}
\begin{center}
\begin{tabular}{c c c c c c c c}
\hline\hline 
$M_\mathrm{1,i}/M_{\odot}$	& $M_\mathrm{2,i}/M_{\odot}$ & $P_{\rm orb,i}/{\rm day}$  & $M_\mathrm{1,f}/M_{\odot}$	& $M_\mathrm{2,f}/M_{\odot}$ & $P_{\rm orb,f}/{\rm day}$ & MT type & Stop Criterion\\
\hline
0.9702 & 0.7686 & 483.94 & 0.4569 & 0.8227 & 1097.3 & Case B & 14 Gyr max age\\
1.0181 & 0.8066 & 303.36 & 0.4327 & 0.8606 & 810.48 & Case B & 14 Gyr max age\\
1.0685 & 0.8465 & 303.36 & 0.4374 & 0.8994 & 861.79 & Case B & donor becomes WD\\
1.1212 & 0.8883 & 190.16 & 0.4156 & 0.9400 & 632.85 & Case B & donor becomes WD\\
1.1766 & 0.9322 & 119.20 & 0.3961 & 0.9827 & 460.23 & Case B & donor becomes WD\\
1.2348 & 0.9782 & 119.20 & 0.3998 & 1.0271 & 489.47 & Case B & donor becomes WD\\
1.2958 & 1.0265 & 190.16 & 0.4274 & 1.0712 & 749.91 & Case B & donor becomes WD\\
1.3598 & 1.0773 & 190.16 & 0.4311 & 1.1176 & 788.89 & Case B & donor becomes WD\\
1.3598 & 1.0773 & 1231.6 & 0.5569 & 1.1036 & 2709.9 & Case C & donor becomes WD\\
1.4270 & 1.1305 & 772.02 & 0.5481 & 1.1542 & 1983.4 & Case C & donor becomes WD\\
\hline
\label{table: selected binaries}
\end{tabular}
\end{center}
\end{table*}

\begin{figure}[tp]
\includegraphics[scale=0.5,angle=0]{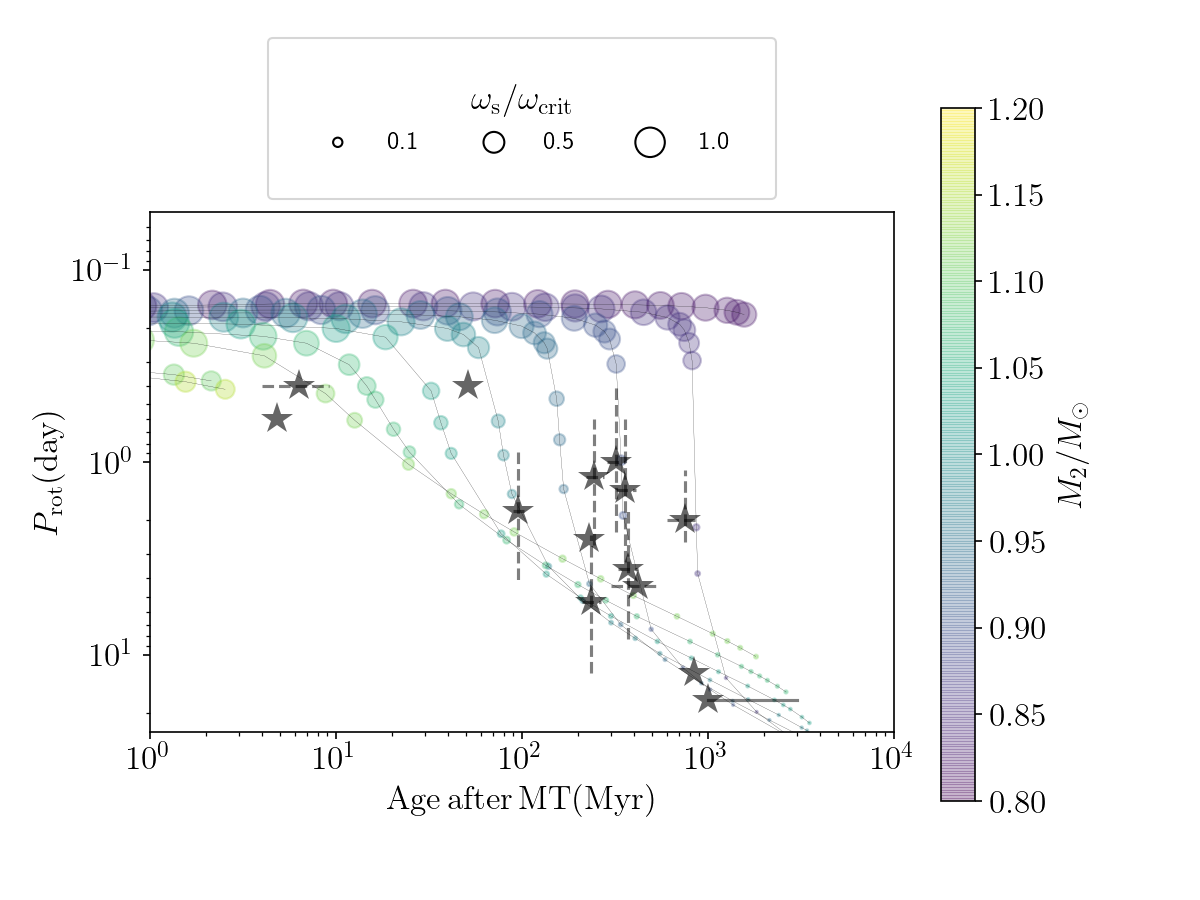}
\caption{Similar to Figure \ref{fig:Prot_age_fiducial_full}, but with only 10 selected post-MT systems described in Table \ref{table: selected binaries}. Lines connect the accretor models from the same evolutionary track.}
\label{fig:Prot_age_selected}
\end{figure}

In Figure \ref{fig:Prot_age_selected}, we present the $P_{\rm rot}$ of the accretor stars from the 10 selected binary tracks as a function of the time elapsed since MT stops. These 10 binaries cover the potential post-MT parameter space, as depicted in Figure \ref{fig:Prot_age_fiducial_full}, in the $P_{\rm rot}$ versus age plane. The data, represented by black stars, still align well with the models. Models from the same binary evolutionary track are connected by thin solid black lines. The two case C evolutionary tracks are the shortest, with the donor star transforming into a WD in roughly a few million years.

The distribution of the models exhibits a mass-dependent preference, with lower-mass accretors (i.e. $M_2<1M_\odot$) slowing down according to G18's magnetic braking prescription over a longer timescale of over 100 Myr. More massive accretors (i.e. $M_2>1.1M_\odot$) begin to decrease in $\omega/\omega_{\rm crit}$ within 10 Myr since MT stops. This phenomenon is associated with the complexity of the star's surface magnetic field, as discussed in detail in Section \ref{subsec:G18 default result} and shown in Figure \ref{fig:Prot_age_model_g18_full_abn_mgnf}. A lower mass star has a deeper convective envelope with a larger $\tau_{\rm conv}$ and (according to the phenomenology proposed by the G18 model) a more complex surface magnetic field. This results in saturated, inefficient magnetic braking, giving these accretor stars a longer spin down time.

\begin{figure}[tp]
\includegraphics[scale=0.5,angle=0]{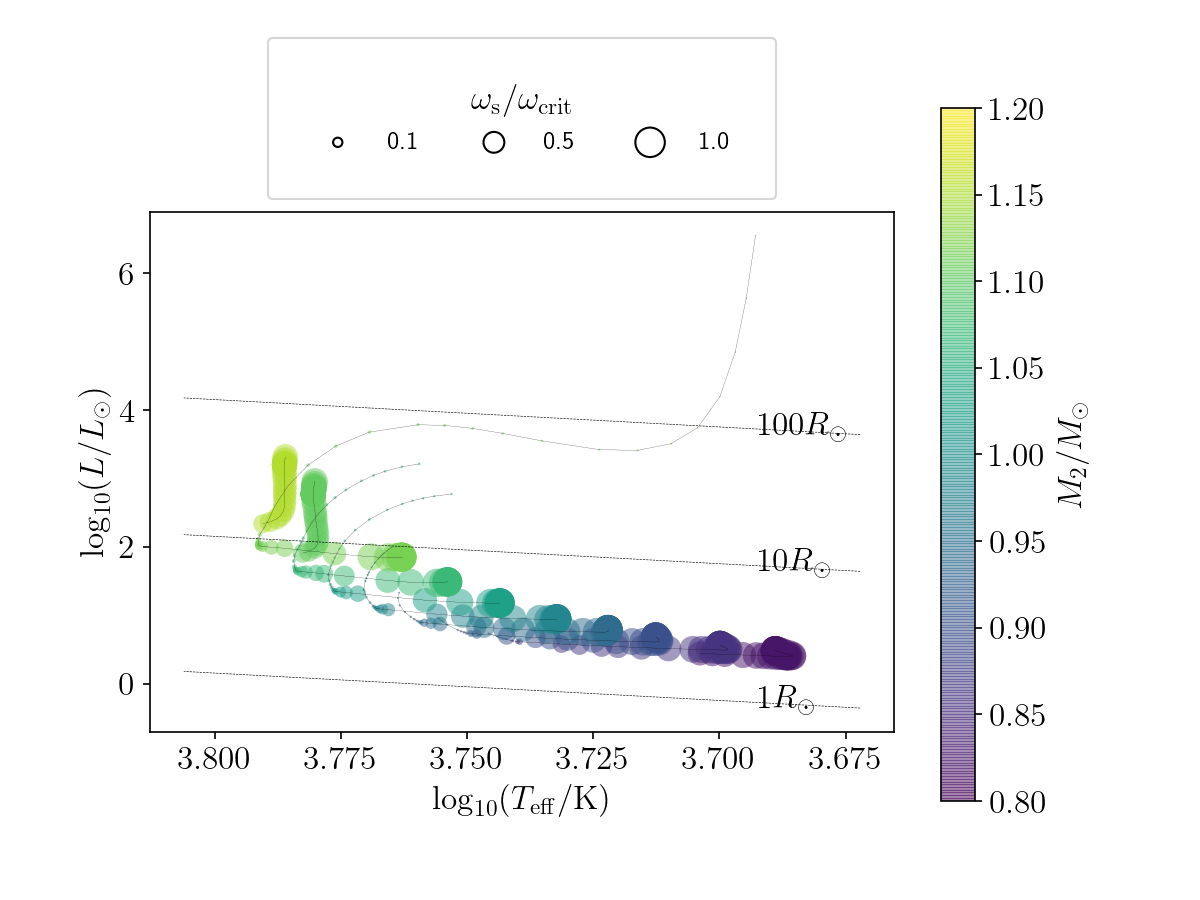} 
\caption{The evolutionary track of the accretor star from Table \ref{table: selected binaries} on the Hertzsprung-Russell diagram. The size and color of the circles follow the same descriptions as in Figure \ref{fig:Prot_age_fiducial_full}.}
\label{fig:HR_selected_binaries}
\end{figure}

Focusing on the same 10 selected binaries from Table \ref{table: selected binaries} and Figure \ref{fig:Prot_age_selected}, we illustrate the evolutionary track of accretor stars after the cessation of MT in the Hertzsprung-Russell diagram in Figure \ref{fig:HR_selected_binaries}. For post case B MT accretors, most stars are still in the early MS evolution right after the accretion, indicated by a central hydrogen mass fraction $X(r=0)$ greater than 0.3. As $X(r=0)$ gradually decreases to 0, the accretor star's evolutionary track shifts towards the upper right direction on the HR diagram. Concurrently, the star's radius increases, decreasing the critical rotation rate and decreasing the surface rotation speed more rapidly, causing the ratio $\omega_s/\omega_{\rm crit}$ to quickly drop below 0.1. Subsequently, a helium core is formed. The parameter regions in Figure \ref{fig:Prot_age_fiducial_full} covered by very small circle size (i.e., where age after MT $>$ 0.2 Gyr and $P_{\rm rot}>3$ days) signify that the spin-down is primarily attributed to stellar expansion due to evolution rather than magnetic braking. For the two case C MT accretors, when their donor stars evolve into WDs then simulation stops, their $X(r=0) > 0.2$, and their evolutionary track does not have a turning point on the HR diagram.

\subsection{A Detailed Analysis of Rotation and Orbital Evolution History}

\begin{figure}[tp]
\includegraphics[scale=0.45,angle=0]{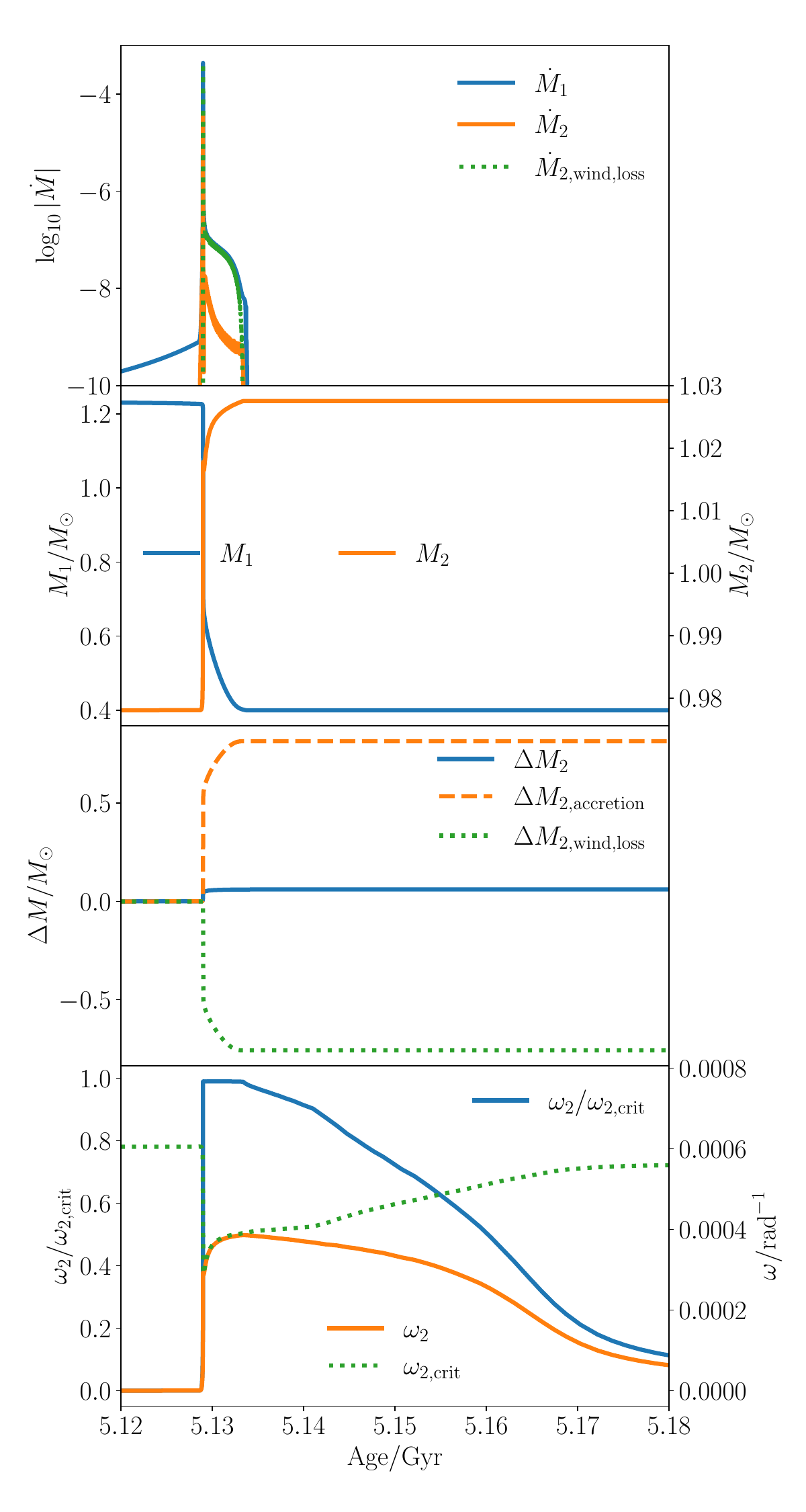} 
\caption{The detailed evolution information for a typical low-mass binary with an initial donor mass of 1.2348 $M_\odot$, accretor mass of $0.9782$ $M_\odot$, and an orbital period of 119.20 days, focusing on the MT phase. The first panel displays the total mass change rate of the donor (blue) and the accretor (orange), along with the mass loss due to the accretor's wind (green dashed). The second panel shows the masses of the donor (blue) and the accretor (orange). The third subplot illustrates the total change in mass of the accretor (blue), the change in mass of the accretor due to RLOF (orange), and the change in mass of the accretor due to wind mass loss (green). The last panel presents the surface rotation speed of the accretor as a fraction of critical rotation (blue), the surface rotation speed of the accretor (orange), and the critical rotation speed of the accretor (green dashed).}
\label{fig:Mdot_t_paper}
\end{figure}

We chose a representative binary system with a solar-type accretor from Table \ref{table: selected binaries}. The initial conditions for this system are $M_{\rm 1,i}=1.2348$ $M_\odot$, $M_{\rm 2,i}=0.9782$ $M_\odot$, and $P_{\rm orb,i}=119.20$ days. This selection allows us to closely examine the evolution of this particular system, with a focus near the MT phase, occurring between the ages of 5.12 and 5.18 Gyr, as shown in Figure \ref{fig:Mdot_t_paper}.

The top panel of Figure \ref{fig:Mdot_t_paper} shows the total mass loss rate of the donor star, $\dot{M}_1$, which includes the mass loss through its own wind and the stable RLOF, occurring when the star fills its Roche-lobe radius. The total mass change rate of the accretor star, $\dot{M}_2$, is the sum of the mass gain from the donor star through the stable RLOF and the mass loss due to its own wind, $\dot{M}_{\rm 2,wind,loss}$. The stable RLOF lasts from 5.129 to 5.135 Gyr. Within this 6 Myr period, because the accretor star quickly reaches its critical rotation rate (see the corresponding bottom panel), its own boosted strong wind is almost balanced by the accretion rate. The $\dot{M}_2$ is two orders of magnitude smaller than $\dot{M}_1$ and $\dot{M}_{\rm 2,wind,loss}$.

In the second panel of Figure \ref{fig:Mdot_t_paper}, we present the mass change history of both the donor and accretor stars. The donor star decreases from 1.23 $M_{\odot}$ to 0.4 $M_\odot$ before and after the stable RLOF. Meanwhile, the accretor's mass increases from 0.98 $M_{\odot}$ to 1.03 $M_{\odot}$. Analyzing from the corresponding third panel, during the stable RLOF, the accretor's mass accumulates to 0.8 $M_{\odot}$, but the boosted wind causes the star to lose 0.75 $M_{\odot}$ of material, resulting in the accretor gaining only 0.05 $M_\odot$ after the cessation of the MT.

In the last panel of Figure \ref{fig:Mdot_t_paper}, we show the rotation evolution of the accretor star. Immediately after the onset of the RLOF, the accretor's rotation speed reaches its critical rotation rate within an order of magnitude of 0.1 Myr. Magnetic braking could not significantly spin down the star within such a small timescale. After the MT, $\omega_s/\omega_{\rm crit}$ begins to drop from 1 to 0.1 mainly due to magnetic braking, over a time period of 43 Myr. After the MT phase, the critical rotation rate of the accretor star increases slightly due to a small decrease in stellar radius, as $\omega_{\rm crit}^2 \propto R^{-3}$.

\section{Other Magnetic Braking Prescriptions}
\label{sec:other MB}

\subsection{M15's Prescription in Slowing Down Stars with a Critical Rotation Rate}

\begin{figure}[tp]
\includegraphics[scale=0.5,angle=0]{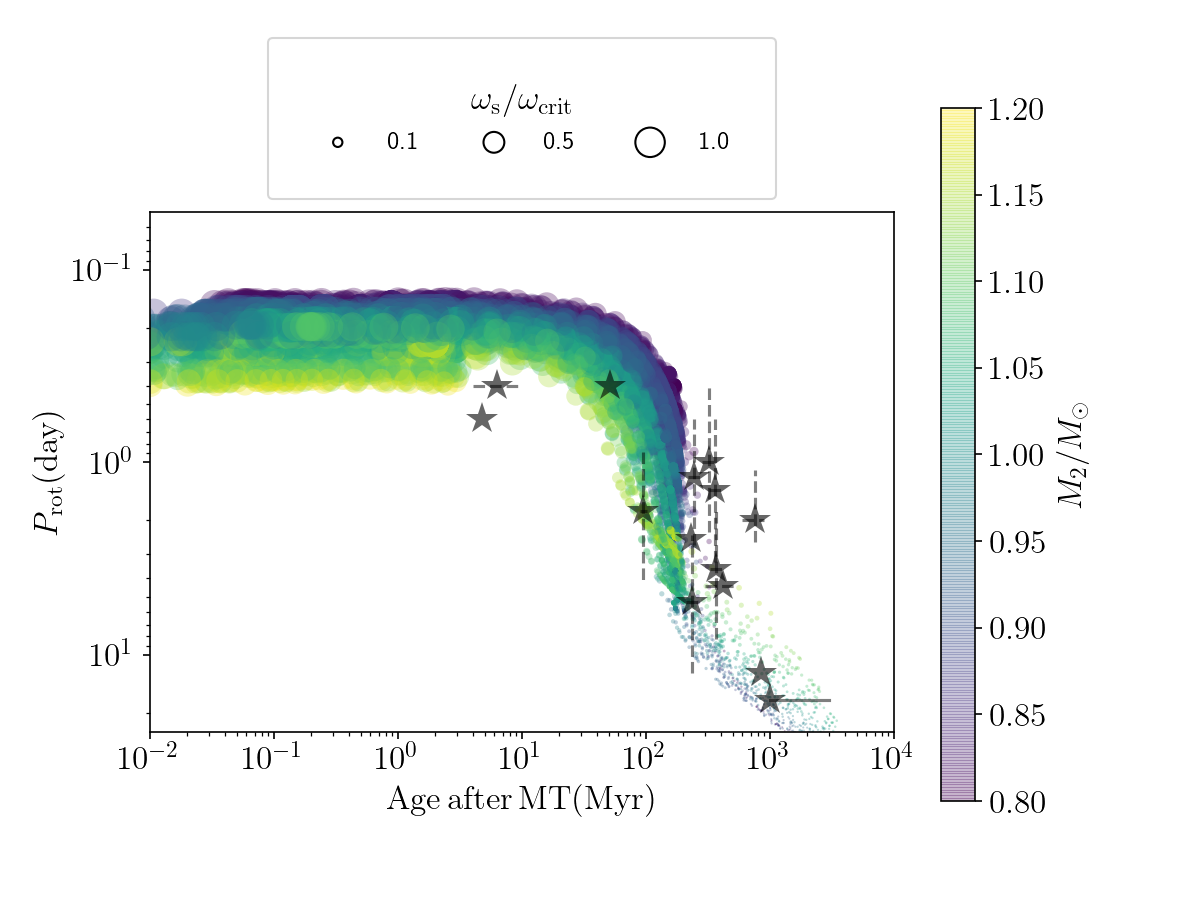}
\caption{This figure follows the description of Figure \ref{fig:Prot_age_fiducial_full} but uses M15's magnetic braking prescription.}
\label{fig:Prot_age_model_consv_m15_full}
\end{figure}

The prescription by \citet{2015ApJ...799L..23M} is developed from 2D MHD simulations to quantify the angular momentum loss $\dot{J}$ of stars due to magnetic braking. In Figure \ref{fig:Prot_age_model_consv_m15_full}, we show the rotation period of the accretor stars as a function of evolutionary time after the stop of the MT. Models that satisfy the five criteria described with bullet points at the beginning of Section \ref{sec:result} are shown on this scatter plot. Regardless of mass, the surface convection zone properties, and the critical rotation period right after the MT, all systems experience the saturated regime with less efficient magnetic braking, where the Rossby number is small ($R_o \ll 1$), and all accretor stars maintain a rapid rotation until $\sim 100$ Myr. After that, magnetic braking reduces the star spin rate significantly in another $\sim 100$ Myr. After $\sim 200$ Myr, most of the accretor stars evolve into a later phase and have $\omega_s/\omega_{\rm crit}< 0.1$, the slowing down in rotation is due to stellar expansion instead of magnetic braking. In general, models applied with M15's prescription are in good agreement with the data, which is represented by black stars in Figure \ref{fig:Prot_age_model_consv_m15_full}. However, three systems (WOCS 4348, WOCS 4230, and WOCS 3001) are slightly outside the region covered by the model.

\begin{figure}[tp]
\includegraphics[scale=0.5,angle=0]{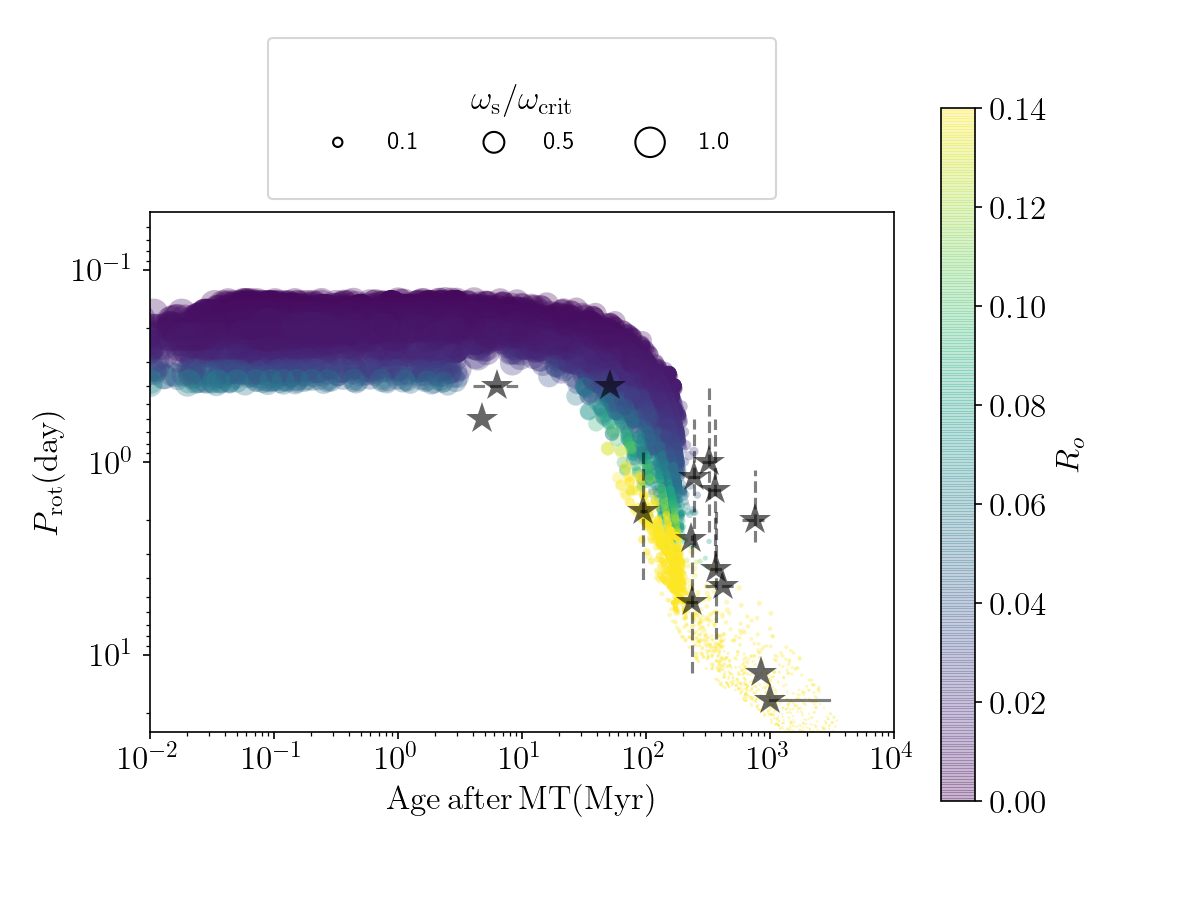}
\caption{This figure follows the description of Figure \ref{fig:Prot_age_model_consv_m15_full}, but the color indicates the Rossby number of the accretor star.}
\label{fig:Prot_age_model_m15_Ro} 
\end{figure}

Under the saturation regime condition, 
\citet{2015ApJ...799L..23M} introduces less efficient magnetic braking if the star's surface Rossby number is smaller than a critical value $R_{\rm o,sat}$ — we applied this value as 0.14 \citep{2019A&A...631A..77A}. Other calibrated parameters are described in Table 1 of \citet{2023ApJ...950...27G}, utilizing solar and open cluster data. In Figure \ref{fig:Prot_age_model_m15_Ro}, we add the Rossby number information as the color variable, indicating the saturation regime occurs before $\sim 100$ Myr for all selected binaries.

\subsection{RVJ's Prescription}

The RVJ prescription \citep{1983ApJ...275..713R} is employed as the default magnetic braking prescription in the \texttt{MESAbinary} module, assuming strong tides resulting in tidal synchronization within the system. Consequently, the binary orbital frequency term from the equation describing how magnetic braking alters the orbits of binaries (see Equation 8 in the third \texttt{MESA} instrument paper \citealt{2015ApJS..220...15P}, which primarily focuses on the release of the binary module), is substituted with the stellar rotation frequency in our format (See Equation 2 in \citealt{2023ApJ...950...27G}). The RVJ prescription originates from the Skumanich law, an empirical relation describing the stellar rotation rate as a function of age. Notably, the RVJ's prescription does not consider the saturation regime, leading to a lack of association between angular momentum loss and surface magnetic field topology, as well as surface convection zone properties. This is the primary distinction between the RVJ's prescription and the other three prescriptions.

\begin{figure}[tp]
\includegraphics[scale=0.5,angle=0]{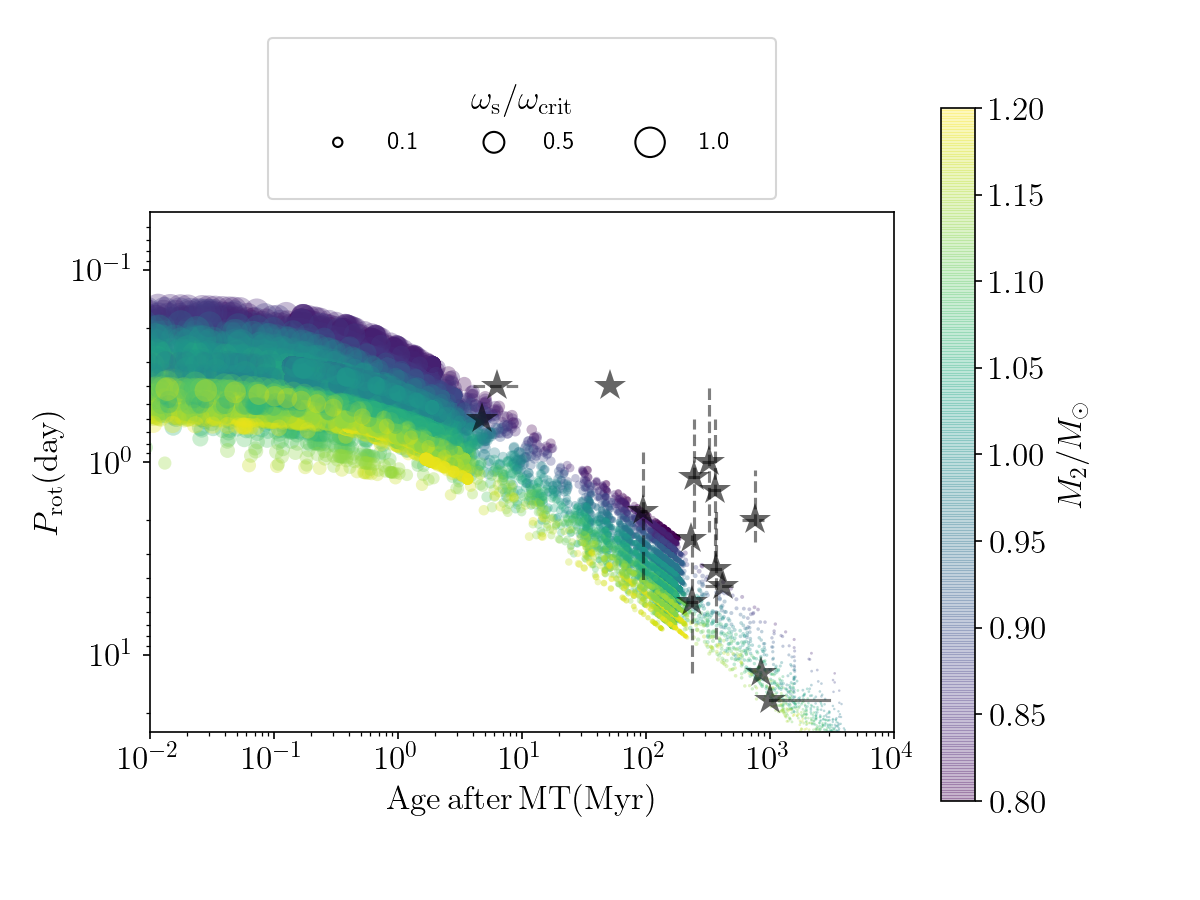}
\caption{This figure follows the description of Figure \ref{fig:Prot_age_fiducial_full} but uses RVJ's magnetic braking prescription.}
\label{fig:Prot_age_model_rvj_full}
\end{figure}

In Figure \ref{fig:Prot_age_model_rvj_full}, we show the rotation period of the accretor star with age since MT stops for the RVJ's prescription. Most of the accretor stars are still near critical rotation right after the cessation of MT. In contrast to the results from G18 and M15's prescriptions, accretor stars experience a significant spin-down due to magnetic braking near 1 Myr since MT stops, which is much earlier than the other two prescriptions. This spin-down applies to all mass ranges from 0.8 to 1.2 $M_\odot$, and the rotation speed drops below 0.5 of its critical rotation. After 100 Myr, the accretor stars experience another spin-down due to stellar expansion as the star evolves to a later stage. From Figure \ref{fig:Prot_age_model_rvj_full}, it is evident that fast-rotating stars (with $P_{\rm rot}\sim 1$ days) between 0.1 to 1 Gyr pose a challenge to the RVJ's prescription.

\subsection{CARB Prescription}

\begin{figure}[tp]
\includegraphics[scale=0.5,angle=0]{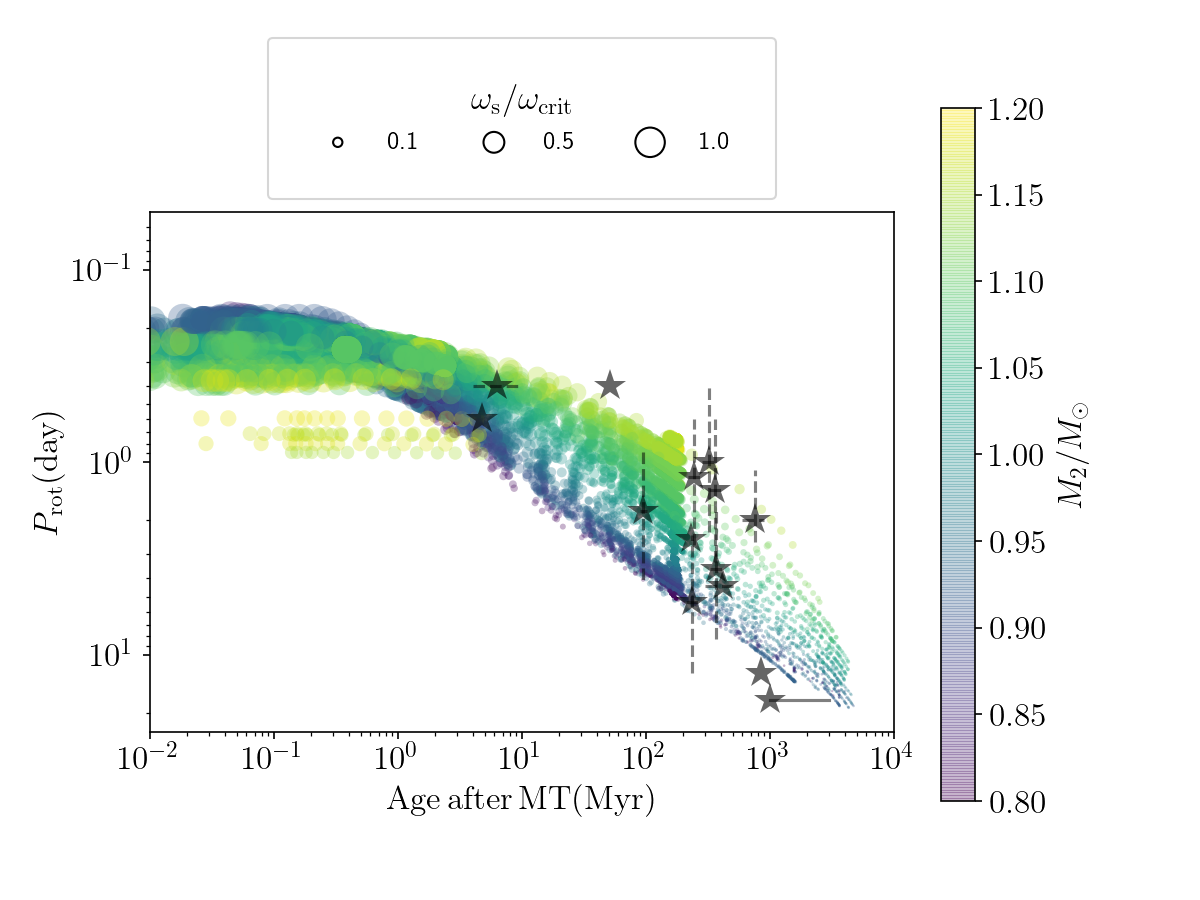}
\caption{This figure follows the description of Figure \ref{fig:Prot_age_fiducial_full} but uses CARB's magnetic braking prescription.}
\label{fig:Prot_age_model_carb_full}
\end{figure}

Similar to RVJ's prescription, CARB's prescription does not take the saturation regime into account. 
In Figure \ref{fig:Prot_age_model_carb_full}, we show the rotation period as a function of age after MT. The color indicates the mass of the accretor stars, and the circle size gives the ratio between the surface and the critical rotation rate. Right after MT, all stars are near a critical rotation rate. Less massive stars (around $0.8$ $M_{\odot}$) begin to reduce the rotation speed at an earlier time, near $\sim 1$ Myr. For more massive stars, where the mass is greater than $1.1$ $M_{\odot}$, the rotation rate drops at a later time, after $\sim 10$ Myr. The reason is that less massive stars have a deeper convective envelope, and the surface eddy turnover time is relatively larger, resulting in a greater angular momentum reduction rate ($\dot{J}\propto {\tau_{\text{conv}}}^{8/3}$). CARB's prescription qualitatively matches the data; only three of the systems do not overlap with the models, with the recalibration performed in \citet{2023ApJ...950...27G}, Section 4.1.4, as shown in their Figure 2. The fast-rotating accretor with a $P_{\rm rot}$ of 0.4 days at 50 Myr might be covered by models with a new set of calibration parameters. The other two slow-rotating data points near 1 Gyr could come from a model with a slowly rotating accretor at the cessation of the MT, with an even more non-conservative  (see more in Section \ref{subsec:less efficient MT}).

\section{Discussion}
\label{sec:Discussion}

\subsection{Less Efficient Mass Transfer Grid Cases}
\label{subsec:less efficient MT}

If the MT efficiency is defined as the mass gain of the accretor star over the mass loss of the donor star, the \textit{rotation limited accretion} is very close to non-conservative MT, where the accretor only gains 2\% to 7\% of its original mass (see Table \ref{table: selected binaries}). This highly non-conservative MT is also observed in other related low-mass binary evolution studies \citep{2023arXiv231107528S}. \texttt{MESA}'s treatment of angular momentum transport during MT follows \citealt{1975ApJ...198..383L,2013ApJ...764..166D}, where the angular momentum could be transferred onto the accretor star either through an accretion disk or by impact with the incoming stream. No matter which mode of accretion is considered, even when we have magnetic braking acting on both stars throughout the entire simulation, the angular momentum transfer during MT is very efficient and can spin up a star to near-critical rotation rates with only a small amount of mass accreted, and within a very short timescale. The boosted wind is balanced by the accretion. The combination of these two processes results in the accretor star gaining only a small amount of material.

\begin{figure}[tp]
\includegraphics[scale=0.5,angle=0]{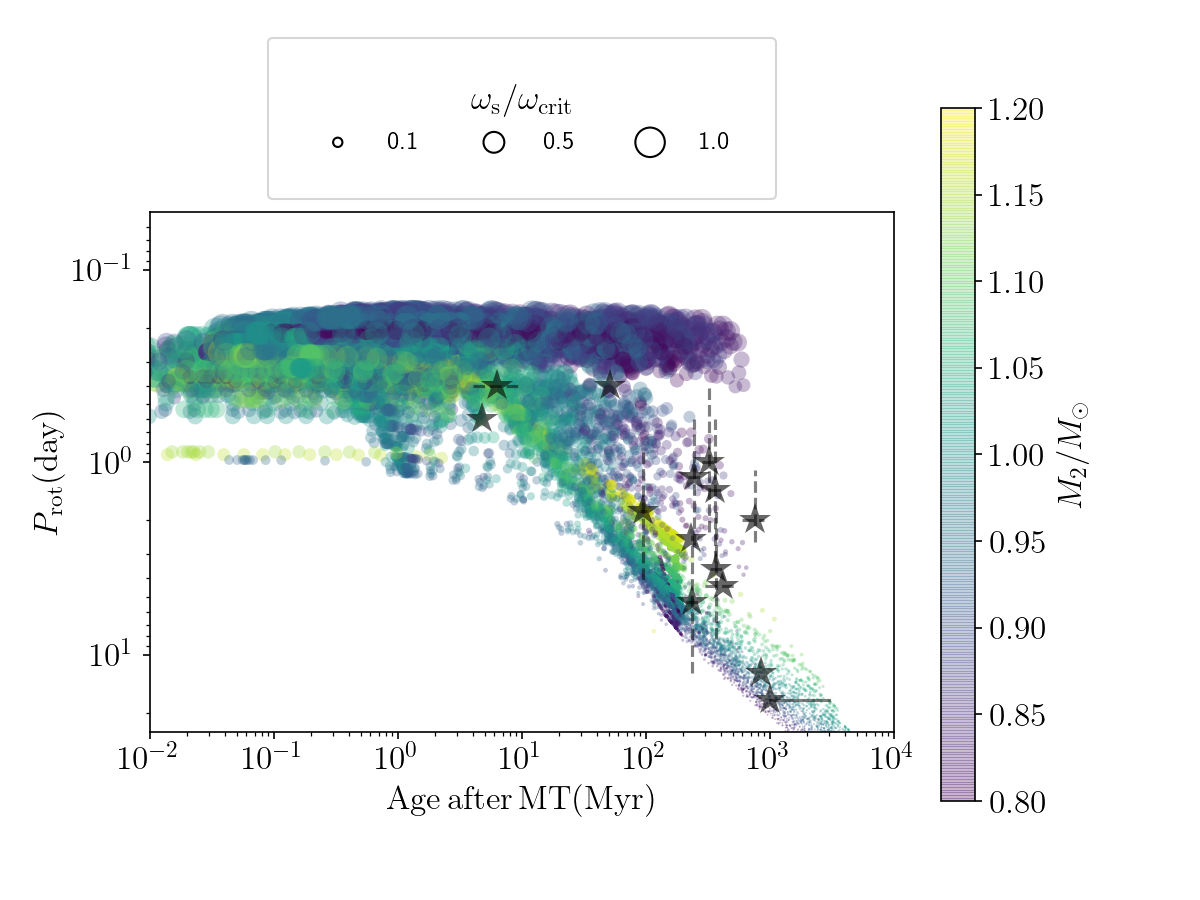}
\caption{This figure follows the description of Figure \ref{fig:Prot_age_fiducial_full} but with a RLOF efficiency $\beta$ of 5\%.}
\label{fig:Prot_age_g18_beta5p_full_1}
\end{figure}

In Figure \ref{fig:Prot_age_g18_beta5p_full_1}, we present another set of models with rotation period as a function of age, but with a non-conservative MT setting. In \texttt{MESA}'s terminology, the RLOF efficiency \texttt{mass\_transfer\_beta}, is set to 0.95, meaning that 5\% of the mass transferred through the RLOF is accreted onto the accretor star. From the perspective of univariate analysis, the \textit{rotation limited accretion} is still applied, wherein stars experience enhanced wind when they reach a critical rotation rate.

The purpose of setting a 5\% MT efficiency is to demonstrate the impact of MT on the post-MT spin rates of our models. Compared with Figure \ref{fig:Prot_age_fiducial_full}, in general, the point size right after the cessation of MT is smaller, indicating sub-critical rotation rates. The models cover a broader range of rotation periods in Figure \ref{fig:Prot_age_fiducial_full} and align well with the data. Specifically, the lower boundary of the model-covered region has shifted downward. For example, in Figure \ref{fig:Prot_age_fiducial_full}, the two youngest and two oldest BSS and WD systems near ages of approximately 10 Myr and 1 Gyr are situated at the edge of the model-covered region. However, in Figure \ref{fig:Prot_age_g18_beta5p_full_1}, they are relatively well matched. In summary, the MT efficiency can have a significant impact on post-MT accretor spin down.

Similar to Figure \ref{fig:Prot_age_fiducial_full}, the models exhibit two branches. The top branch experiences a saturation regime and is dominated by low-mass stars with masses below 0.9 $M_{\odot}$, where the surface has a larger convective turnover time ($\tau_{\rm conv}$). The lower branch is dominated by stars with masses greater than 1 $M_{\odot}$, where a significant spin-down is observed near 10 Myr.

\subsection{Performance of Magnetic Braking Prescriptions}

In \citet{2023ApJ...950...27G}, the models are implemented with the same four magnetic braking prescriptions as in this manuscript. A series of single star evolution models is compared with rotation measurements from four young open clusters. Although none of the models could explain the stalled spin-down near 0.7 to 1 Gyr with star mass $<$ 0.8 $M_{\odot}$, the G18 and M15 prescriptions generally agree well with the data. Part of the reason is that both prescriptions take the saturated regime into account. In the comparison of ultracompact X-ray and low-mass X-ray binaries, where systems range from approximately 20 minutes to 12 days in orbital period, systems are assumed to be tidally synchronized. The G18, CARB, and RVJ prescriptions generally provide a better match to the observed orbital period distribution. Furthermore, the orbital period distribution of Cataclysmic Variables favors the G18 model, which takes into account magnetic field complexity \citep{2018ApJ...868...60G,2022MNRAS.517.4916E}.

In this work, among all 14 data points, G18's model is the most promising one to explain the spin-down of all the young and old post-MT accretors, although several studies call into question the degree to which low mass stars exhibit high magnetic field complexity \citep{2018ApJ...854...78F,2019ApJ...876...44F,2019ApJ...876..118S,2019ApJ...886..120S,2020ApJ...894...69S}. CARB's prescription is another promising model to explain the data, except for RE 0044+09 with a post-MT age of 51 Myr and $P_{\rm rot}$ of 0.4 day. Although KOI-3278 (post-MT age 663 Myr with $P_{\rm rot}$ of 12.5 days) and KIC 6233093 (post-MT age $>$ 1 Gyr with $P_{\rm rot}$ of 17.1 days) do not overlap with CARB's model in Figure \ref{fig:Prot_age_model_carb_full}, they could be explained by non-critical rotation models right after MT stops. As discussed in Section \ref{subsec:less efficient MT}, starting with non-critical rotation models through a less efficient MT, the lower boundary of the models shifts downward in the $P_{\rm rot}$ versus age plane. M15's model is a little bit away from the data between 320 to 750 Myr in WD cooling age and 1 to 2 days in $P_{\rm rot}$. RVJ's models are too far away from RE 0044+09, WOCS 4348 (post-MT age of 245 Myr and $P_{\rm rot}$ of 1.2 days), WOCS 4230 (post-MT age of 320 Myr and $P_{\rm rot}$ of 1.0 day), WOCS 2679 (post-MT age of 360 Myr and $P_{\rm rot}$ of 1.4 days), and WOCS 3001 (post-MT age of 750 Myr and $P_{\rm rot}$ of 2 days).

\subsection{Angular Momentum Transfer During the Mass Transfer}

Our main findings are based on angular momentum transport along with , as described by \citealt{1975ApJ...198..383L,2013ApJ...764..166D}. An enhanced wind turns on when the accretor star rotates at its critical rotation rate \citep{1997ASPC..120...83L,1998A&A...329..551L}, while stars maintaining slightly slower rotation rates than the critical rate. However, how much material the accretor star can accept and how angular momentum is transferred remain open questions in binary evolution. 

Instead of an enhanced wind which is more suitable for massive stars, other works propose that viscous processes could transport angular momentum outward through the star and the accretion disk, allowing stars to continue accreting mass without significant angular momentum transfer to avoid rotating at the critical rate \citep{1991ApJ...370..597P,1991ApJ...370..604P}. In other works, it is speculated that at critical rotation, a decretion disk may develop, suggesting that the accretor could continue to accumulate mass even while rotating at critical rates. However, the detailed evolution of such systems remains largely unexplored. Following the approach outlined by \citet{2023MNRAS.519.1409L}, an equatorial circumbinary outflow forms at high  rates, increasing the likelihood of L2 mass loss, particularly under superthermal accretion conditions. Additionally, for the enhanced wind prescription proposed by \citep{1998A&A...334..210H}, the wind is driven by radiation pressure in massive stars. In low-mass stars, if radiation is not strong enough to drive the wind, material around the accretor could accumulate until it fills its own Roche lobe and forms a contact binary. Investigating contact binaries represents another direction for future research, although current limitations exist in modeling these systems using one-dimensional codes.

From the case studies mentioned in Table \ref{table: selected binaries}, lower mass stars could obtain more mass, up to 7\% of their initial mass. Other detailed case studies of blue stragglers \citep{2021ApJ...908....7S, 2023ApJ...944...89S} indicate that blue stragglers gain 20\% to 30\% of their initial mass without using the \textit{rotation limited accretion}. One future direction is to still allow both mass and angular momentum transfer onto the accretor star. However, once it has been accelerated to near critical rotation, angular momentum transfer is significantly reduced to allow more mass to be accreted. This approach could provide a better understanding of the mass distribution of post-MT systems, as well as their structure and surface magnetic field associated with magnetic braking.

\section{Conclusions}
\label{conclusion}

As an extension research based on \citet{2023ApJ...950...27G}, this work no longer uses the strong tide assumption approach by examining four magnetic braking prescriptions in \texttt{MESA} binary simulations. In Figure 1 of \citealt{2023ApJ...950...27G}, with an initial rotation rate ranging from 0.02 to 0.15 of its critical rotation rate, the model overlaps well with the observational single star rotation data from young open clusters. In our model, during MT, the angular momentum transfer is very efficient and can spin up the accretor star to its critical rotation rate. Therefore, studying magnetic braking in binary evolution could help test the magnetic braking prescriptions under extreme conditions: critically rotating stars.


We survey the parameter space of low-mass stars in binary systems, focusing on models that have undergone stable MT. Our findings demonstrate that all magnetic braking prescriptions are insufficient to prevent accretor stars in low-mass binaries from reaching critical rotation rates during the MT phase. After MT, the donor star has already evolved into a WD, while the accretor star remains on the MS. In accordance with the data, the systems after MT have an orbital period ranging from 100 to 5000 days are selected for analysis. Because of the wide orbital separation, tides are weak, and the stellar spins are not synchronized with the orbit. Studying stellar spin-down in these post-MT systems could be approached similarly to single stars in open clusters. Moreover, magnetic braking has no impact on changing the system's orbital separation after MT.

Our simulations show that the G18 magnetic braking prescription better matches the observations. The models derived from G18's prescription have two branches, where this separation could be explained by the complexity of the surface magnetic field. The stellar surface dominated by a dipole magnetic field can be efficiently spun down by magnetic braking. Higher-order magnetic fields result in less efficient spin-down, so there are models that still have high rotation speeds after MT. If the mass or surface information of the stars can be acquired from observation, those post-MT accretor stars could serve for gyrochronology to determine the age since MT stopped.

Beyond approximately 200 Myrs since the cessation of MT, accretor in our grid grow in radius and spin down. The resulting combination of magnetic braking (early) and stellar evolution (later) results in final $\omega_s/\omega_{\rm crit}$ values $<$ 0.1.

Models applied with RVJ and CARB's prescriptions start to spin down at an earlier time, as early as 1 Myr after MT stops. In contrast, for G18 and M15's prescriptions, after 10 Myr, accretor stars could experience spin down due to magnetic braking. The reason is that G18 and M15's prescriptions consider magnetic field saturation, but RVJ and CARB do not. The saturation regime leads to less efficient spin-down for critically rotating stars.


As we only have 14 post-MT systems with rotation measurements, we suspect that the data contain selection effects. For instance, most of the systems have a hot WD (and therefore we assume have only recently completed the MT phase). The flux of the WDs decreases dramatically during the cooling phase, which challenges us in obtaining the UV spectroscopy of those older and colder WDs. Therefore, we cannot perform a population study based on these 14 data points. More observational data would help us better constrain the angular momentum evolution modeling in binary evolution. Finally, we note that the new magnetic braking models (e.g., the recent results for fully convective low-mass stars by \citealt{2023MNRAS.526..870S} and its extension to solar-type stars by \citealt{2024arXiv240205912S}) can be implemented in our code for future investigations.

\section*{Acknowledgments}

We greatly appreciate the referee's valuable insights. M.S. thanks the support from the GBMF8477 grant (PI: Vicky Kalogera) and the extensive support provided by Vicky Kalogera in every aspect. M.S. extends appreciation to the entire \texttt{POSYDON} developer team for their invaluable technical assistance. Special thanks are given to Steve Lubow for offering insightful comments on angular momentum transfer in binaries with magnetic braking. Chenliang Huang is acknowledged by M.S. for engaging discussions that spanned from the inception of the idea to the final completion of this manuscript. S.G. acknowledges the funding support by a CIERA Postdoctoral Fellowship. A.M.G. appreciates the support from the NSF AAG grant No. AST-2107738. E.M.L. received partial support through a CIERA Postdoctoral Fellowship. 

\software{\texttt{numpy} \citep{2020NumPy-Array}, \texttt{scipy} \citep{2020SciPy-NMeth}, \texttt{pandas} \citep{mckinney2010data}, \texttt{Matplotlib} \citep{Hunter2007}, \texttt{astropy} \citep{2013A&A...558A..33A,2018AJ....156..123A}, \texttt{MESA} 
\citep{2011ApJS..192....3P,2013ApJS..208....4P,2015ApJS..220...15P,2018ApJS..234...34P,2019ApJS..243...10P,2023ApJS..265...15J}, \texttt{POSYDON} \citep{2023ApJS..264...45F}.}

\bibliography{rot_references.bib}{}
\bibliographystyle{aasjournal}
\end{CJK*}
\end{document}